%% file: sample-sigconf.tex
\documentclass[sigconf]{acmart}
\copyrightyear{2024}
\acmYear{2024}
\setcopyright{acmlicensed}
\acmConference[WWW '24 Companion] {Companion Proceedings of the ACM Web Conference 2024}{May 13--17, 2024}{Singapore, Singapore.}
\acmBooktitle{Companion Proceedings of the ACM Web Conference 2024 (WWW '24 Companion), May 13--17, 2024, Singapore, Singapore}
\acmISBN{979-8-4007-0172-6/24/05}
\acmDOI{10.1145/3589335.3648298}
\usepackage{multirow}
\usepackage{graphicx}
\usepackage{makecell}
\usepackage{bbm}
\usepackage{amsmath}
\usepackage{xspace}

\AtBeginDocument{%
  \providecommand\BibTeX{{%
    \normalfont B\kern-0.5em{\scshape i\kern-0.25em b}\kern-0.8em\TeX}}}

\begin{document}

\title{Large Language Model based Long-tail Query Rewriting in Taobao Search}

\author{Wenjun Peng}
\authornote{This work was done when the first author was an intern at Taobao Main Search.}
\email{pengwj@mail.ustc.edu.cn}
\affiliation{%
  \institution{University of Science and Technology of China \& State Key Laboratory of Cognitive Intelligence}
  \city{Hefei}
  \state{Anhui}
  \country{China}
}

\author{Guiyang Li}
\email{liguiyang.lgy@taobao.com}
\affiliation{%
  \institution{Taobao and Tmall Group}
  \city{Hangzhou}
  \state{Zhejiang}
  \country{China}
}

\author{Yue Jiang}
\email{jy270069@alibaba-inc.com}
\affiliation{%
  \institution{Taobao and Tmall Group}
  \city{Hangzhou}
  \state{Zhejiang}
  \country{China}
}

\author{Zilong Wang}
\email{huanshi.wzl@taobao.com}
\affiliation{%
  \institution{Taobao and Tmall Group}
  \city{Hangzhou}
  \state{Zhejiang}
  \country{China}
}

\author{Dan Ou}
\authornote{Corresponding author 1}
\email{oudan.od@taobao.com}
\affiliation{%
  \institution{Taobao and Tmall Group}
  \city{Hangzhou}
  \state{Zhejiang}
  \country{China}
}

\author{Xiaoyi Zeng}
\email{yuanhan@taobao.com}
\affiliation{%
  \institution{Taobao and Tmall Group}
  \city{Hangzhou}
  \state{Zhejiang}
  \country{China}
}

\author{Derong Xu}
\email{derongxu@mail.ustc.edu.cn}
\affiliation{%
  \institution{University of Science and Technology of China \& State Key Laboratory of Cognitive Intelligence}
  \city{Hefei}
  \state{Anhui}
  \country{China}
}

\author{Tong Xu}
\authornote{Corresponding author 2}
\email{tongxu@ustc.edu.cn}
\affiliation{%
  \institution{University of Science and Technology of China \& State Key Laboratory of Cognitive Intelligence}
  \city{Hefei}
  \state{Anhui}
  \country{China}
}

\author{Enhong Chen}
\email{cheneh@ustc.edu.cn}
\affiliation{%
  \institution{University of Science and Technology of China \& State Key Laboratory of Cognitive Intelligence}
  \city{Hefei}
  \state{Anhui}
  \country{China}
}

\renewcommand{\shortauthors}{Peng, et al.}

\input{src/abstract}

\ccsdesc[500]{Information systems~Query reformulation}
\ccsdesc[500]{Computing methodologies~Natural language processing}

\keywords{Query reformulation; large language models; semantic matching}



\newcommand{\method}{BEQUE\xspace}
\newcommand{\nothing}{few-recall\xspace}
\maketitle
\input{src/introduction}

\input{src/relatedworks}

\input{src/method}
\input{src/experiments}
\input{src/conclusion}
\bibliographystyle{ACM-Reference-Format}
\bibliography{sample-sigconf}

\appendix

\end{document}

%% file: src/abstract.tex
\begin{abstract}
In the realm of e-commerce search, the significance of semantic matching cannot be overstated, as it directly impacts both user experience and company revenue. 
Along this line, query rewriting, serving as an important technique to bridge the semantic gaps inherent in the semantic matching process, has attached wide attention from the industry and academia.
However, existing query rewriting methods often struggle to effectively optimize long-tail queries and alleviate the phenomenon of \textit{``\nothing''} caused by semantic gap.
In this paper, we present \textbf{\method}, a comprehensive framework that \textbf{B}ridges the s\textbf{E}mantic gap for long-tail \textbf{QUE}ries. 
In detail, \method comprises three stages: multi-instruction supervised fine tuning~(SFT), offline feedback, and objective alignment. 
We first construct a rewriting dataset based on rejection sampling and auxiliary tasks mixing to fine-tune our large language model (LLM) in a supervised fashion.
Subsequently, with the well-trained LLM, we employ beam search to generate multiple candidate rewrites, and feed them into Taobao offline system to obtain the partial order. 
Leveraging the partial order of rewrites, we introduce a contrastive learning method to highlight the distinctions between rewrites, and align the model with the Taobao online objectives.
Offline experiments prove the effectiveness of our method in bridging semantic gap. 
Online A/B tests reveal that our method can significantly boost gross merchandise volume (GMV), number of transaction (\#Trans) and unique visitor (UV) for long-tail queries.
\method has been deployed on Taobao, one of most popular online shopping platforms in China, since October 2023. 

\end{abstract}

%% file: src/introduction.tex
\begin{figure}[!t]
  \centering
  \includegraphics[width=0.4\textwidth]{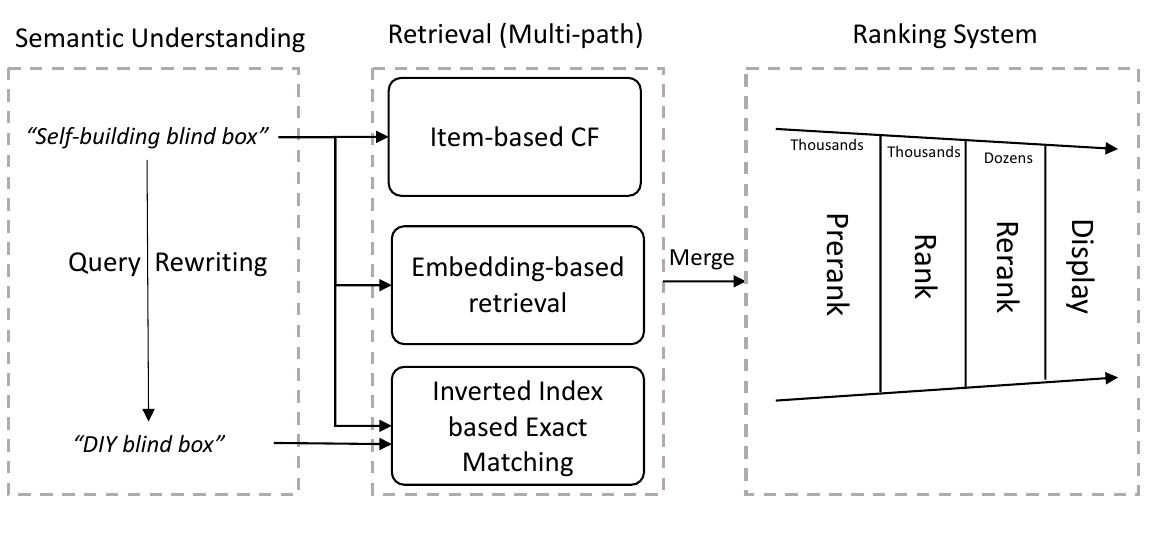}
  \caption{Framework of Taobao search engine.}
  \label{fig:taobao-framework}
\end{figure}

\section{Introduction}
Past decades have witnessed the exceptionally rapid growth of e-commerce platforms. Leading e-commerce companies, such as Taobao, JD and Amazon, have amassed hundreds of millions of users, generating billions of gross merchandise volume (GMV) annually. To facilitate the quick retrieval of related products for these users, a well-established search paradigm has been proposed, as illustrated in Figure~\ref{fig:taobao-framework}, Specifically, this paradigm involves several steps, i.e., \textit{``semantic understanding - retrieval - rank''}. 
Among them, \emph{semantic understanding} serves as the foundation of entire system, ensuring accurate matching of user intent. However, due to the variations for how users express their preferences for products, semantic gaps often exist between their queries and the product keywords, even worse with long-tail queries where retrieval system may fail to provide any relevant products. For instance, a user with personal expression habits may input a long-tail query like \textit{"self-building blind box"}, which will lead to more retrieval results if with its synonymous query like \textit{"DIY blind box"}. Unfortunately, traditional term-matching solutions like inverted index could probably fail to match the commonly-used \textit{``DIY''} with non-customary term \textit{``self-building''}, which limit the retrieval results and significantly impair the user experience. Therefore, it is urgently required to solve the semantic gap challenge for long-tail queries, and address the problem of \textit{``\nothing''} in e-commerce platform.

Traditionally, prior arts~\cite{huang2020embedding, li2021embedding, zhang2020towards, wang2023learning} mainly focus on the \textit{``embedding based retrieval''} paradigm, which initially map the queries and products into a common semantic space, and then support the Top-$K$ retrieval with approximate nearest neighbor (ANN) methods. 
However, the retrieval outcomes might be difficult to interpret, which severely limit the performance. To enhance the controllability of retrieval outcomes, some efforts have been made on the \textit{``query rewriting \& exact match''} paradigm. On the one hand, the \emph{discriminative methods}~\cite{zheng2020bert, li2022query} attempt to ``rewrite” queries via finding similar terms from a query reformulation set, and then utilize them to search for relevant products using sparse retrieval. 
Although these approaches could effectively expand the semantic of hot queries, long-tail queries may not be adequately optimized, thus no related rewrite can be generated. 
On the other hand, the \emph{generative methods}~\cite{qiu2021query, wang2021queen} involve supervised training on \textit{<query, rewrite>} pair data to empower the model with rewriting capabilities, and the alignment process~\cite{mohankumar2021diversity, Agrawal2023} is further incorporated to enhance the metric preference. 
Although these methods partially address the semantic gap problem, they typically rely on small generative models with limited comprehension of long-tail queries, which significantly constrained the rewriting capability. 
Recently, with the development of LLM techniques, some efforts~\cite{wang2023query2doc, jagerman2023query, anand2023query, wang2023can} solely utilize LLMs as retrieval data augmentation generators without additional training to expand query semantics. 
However, these methods, even with carefully curated prompts, may still constrain the ability to specialize for query rewriting task, leading to the poor alignment with objectives of e-commerce search.

To effectively bridge the semantic gap for long-tail queries, and solve the above challenges via producing controlled and aligned outcomes with integrating the knowledge of LLMs, we propose \method, a novel framework that involves three stages of fine-tuning LLMs, namely \emph{multi-instruction supervised fine-tuning} (SFT), \emph{offline system feedback}, and \emph{objective alignment}. 
Specifically, in the first stage, we utilize the rejection sampling to collect \textit{<query, rewrite>} pairs with desired quality distribution, and then combine these pairs with data from quality classification, product title prediction and chain of thought (CoT) tasks to construct the multi-instruction rewriting dataset for fine-tuning our LLM.
Next, with the well-trained LLM, we employ beam search to generate multiple candidate rewrites for each query. 
These candidate rewrites are fed into the Taobao offline system to retrieve a collection of related products. 
We calculate the quality score of retrieved products for the rewrites and use them as rewards for candidates ranking. To calibrate generation probability of the candidate rewrites, we introduce a Bradley-Terry based contrastive learning method that considers the partial order among the these rewrites. Ultimately, the model training objective is aligned with the online goal of the Taobao search, ensuring that the generated rewrites yield the desired search results.

The main contributions of this work are listed as follows:

\begin{itemize}
\item  We have analyzed long-tail queries in e-commerce search and identified the semantic gap problem associated with such queries. To the best of our knowledge, we are the first to fine-tune LLMs for industrial query rewriting task.
\item We propose a three-stage fine-tuned framework called \method to address the issue of semantic gap in long-tail queries. This framework is designed to generate rewrites that align with the objectives of Taobao search.
\item The effectiveness of our model is demonstrated through both offline and online experiments, showcasing its ability to significantly improve e-commerce revenue.
\end{itemize}

%% file: src/relatedworks.tex
\section{Related Works}
\subsection{Query Rewriting}
Query rewriting, also known as query expansion or query reformulation, plays a pivotal role in e-commerce search technology and has a profound impact on the user's shopping experience and the revenue of e-commerce platforms. 
This technique can be broadly categorized into discriminative and generative methods.

\textbf{Discriminative methods} treat query rewriting as a retrieval process that expand the semantics of the original query by selecting appropriate terms from the candidate set. 
For example, pseudo-relevance~\cite{cao2008selecting, tao2006regularized, xu2017quary} selects the top k documents from the initial retrieval as semantic extensions. 
These approaches, however, often pose challenges in effectively controlling the semantic scope and ensuring retrieval relevance.
To address these challenges, one potential solution is to utilize a well-built thesaurus~\cite{mandal2019query, bhogal2007review} as a candidate rewrite set.
However, it is important to note that the effectiveness of these methods highly depends on the quality of the thesaurus. 
Inadequate quality may result in query semantic drift, where the intended meaning of the query is compromised.
Furthermore, alternative approaches~\cite{cui2002probabilistic, antonellis2008simrank++, manchanda2019intent, li2022query} involve generating candidate rewrites based on search logs, incorporating similar terms from users' search history as extensions.
Unfortunately, due to the Matthew effect, search logs naturally exhibit a bias towards popular queries, resulting in that the training data collected through this approach may not sufficiently meet the optimization needs for less frequently searched long-tail queries.


\textbf{Generative methods}~\cite{lee2018rare, qiu2021query, song2021triangular, wang2021queen} formulate the rewriting task as a generative process, where a transformer-like model is utilized to generate candidate terms for the original query.
Some methods~\cite{Agrawal2023, mohankumar2021diversity} incorporate reinforcement learning or contrastive learning to align with human preferences or offline metrics. 
These methods have the ability to generate related rewrites for each query, but usually employ a model with limited number of parameters, which makes them less effective in processing long-tail queries that require a deeper level of semantic understanding. Besides, their generated rewrites are often inconsistent with the optimization goals of the actual search engine. 
On the other hand, LLM-based rewriting methods~\cite{wang2023query2doc, jagerman2023query, anand2023query, wang2023can} provide a deeper understanding and can generate appropriate expansion for long-tail queries. 
Nevertheless they do not receive ad-hoc training on fine-tuning and goal alignment, and lack of specialization and   controllability to the query rewriting task, which may potentially introduce illusions and noise to the original query.

\subsection{Preference Alignment}
In recent years, with the increase in number of parameters, language models~\cite{ouyang2022training, touvron2023llama, du2022glm, yang2023baichuan} have demonstrated incredible semantic understanding and zero-shot capabilities~\cite{peng2023gpt, xu2023large} on one hand. 
However, they have also faced the challenges of model hallucinations and ethical issues~\cite{ouyang2022training, zhang2023siren, lyu2024crud}. 
These models have the potential to fabricate facts with their extensive background knowledge~\cite{wang2023survey}, resulting in misleading to users.
To align the model's outputs with human morals and preferences, reinforcement learning~(RL) has been introduced to force the model to learn the partial order among different outputs~\cite{stiennon2020learning, christiano2017deep}. 
For instance, OpenAI made a groundbreaking application~\cite{ouyang2022training} of RL to the training process of large language models, specifically with ChatGPT, which received tremendous attention.
Besides, LLama2~\cite{touvron2023llama} designed a multi-objective reward function that not only ensures the safety of model outputs but also enhances their helpfulness. 
Unfortunately, these methods often rely on complex training processes, and abundant high-quality data, making them difficult to tune with vast number of hyperparameters.
Therefore, rejection sampling-like methods~\cite{deng2020residual, bai2022constitutional, gulcehre2023reinforced, dong2023raft} have been proposed to continue training the model by collecting outputs with high rewards from the previous rounds, aligning LLMs with human preferences.
Moreover, contrastive learning methods~\cite{song2023preference, yuan2023rrhf, rafailov2023direct} directly rank the outputs based on their rewards and utilize ranking loss to adjust the output probabilities, explicitly learning the partial order of outputs. 
Building upon previous findings, we further propose three custom-designed Taobao metrics as ranking rewards to calibrate generation probabilities of candidate rewrites, forcing model to align with online objectives of Taobao search.

%% file: src/method.tex
\begin{figure*}[!t]
  \centering
  \includegraphics[width=0.82\textwidth]{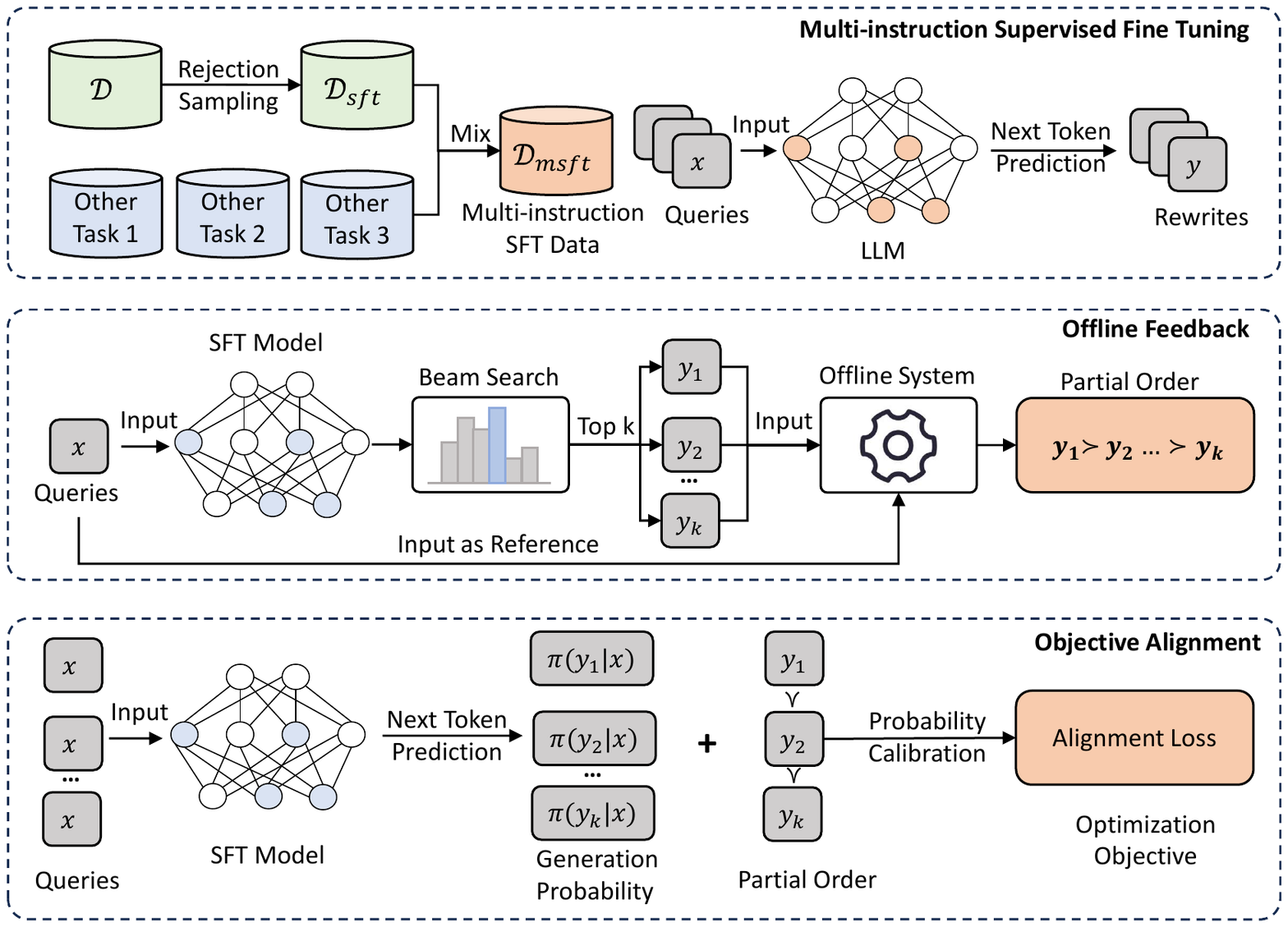}
  \caption{Framework of \method.}
  \label{fig:LLMs4QR-framework}
\end{figure*}

\section{Method}

\subsection{Framework Overview}
Long-tail query rewriting aims to expand the original query semantics to address the problem of semantic gap while ensuring relevance.
To this end, as shown in Figure~\ref{fig:LLMs4QR-framework}, we propose a three-stage rewriting framework, which consists of: multi-instructions supervised fine-tuning (SFT), offline feedback and objective alignment.
1) First, with rejeciton sampling, we constructed a multi-instructions SFT dataset based on online logs that focuses on rewriting tasks, mixed with quality classification, query correction and chain of thought~(CoT) tasks to train rewriting-specific LLMs.
2) After that, we use the well-trained LLM obtained in the first stage to generate multiple candidate rewrites for each sampled query. 
In order to obtain the partial order of these candidate rewrites, we construct a taobao offline system to obtain search results for these rewrites. 
The quality scores of the search results are used to rank the candidates.
3) Based on the partial order of candidate rewrites, we calibrate the generation probability of these rewrites using Bradley-Terry based contrastive learning to maximize the probability of rewrites that can obtain the desired search results.

\subsection{Multi-instruction SFT}
\label{sec:MSFT}
Given that no publicly available LLMs are specifically designed for e-commerce query rewriting, direct utilization of general LLMs to address the long-tail query semantic gap issue is likely to introduce inaccuracies and noise. 
Consequently, we have embarked on an approach wherein we gather various rewriting-related tasks to fine tune LLMs, enhancing their ability to comprehend and rewrite e-commerce queries effectively.

\noindent \textbf{Query Rewriting Dataset}: we initially source rewrites from Taobao previous-generation rewriting policy. This process yields the initial rewriting dataset. 
Specifically, when a user initiates a query $x$ in Taobao search, old rewriting policy generates a list of rewrites queries $Y=\{y_1, y_2, \cdots, y_n\}$. 
From this list, we select the top-ranked $y_1$ as the gold standard candidate to construct our initial rewriting dataset $\mathcal{D}$ with $N$ samples:
\begin{equation}
\mathcal{D}=\left\{\left.\left(x^i, y^i\right)\right|_{i=1} ^N \text { such that } x^i \sim p(x), y^i \sim \pi_{old}\left(y \mid x=x^i, \theta_{old}\right)\right\},
\end{equation}
where $p(x)$ denotes the query distribution in Taobao search engine, $\pi_{old}$ and $\theta_{old}$ is the previous-generation rewriting policy of Taobao and it's parameters.

It's important to highlight that e-commerce query rewriting differs from other text generation tasks. 
In this context, semantic similarity between query and rewrite does not necessarily guarantee retrieval of similar sets of products. 
What we aim to achieve is a high relevance between the products retrieved by rewrite $y$ and the original query $x$. 
To attain this, we apply a relevance filter to $\mathcal{D}$ through rejection sampling:

\begin{equation}
\mathcal{D}_r=\left\{\left.\left(x^i, y^i\right)\right|_{i=1} ^ {N_r} \text { such that } x^i, y^i \in \mathcal{D}, rele(x^i, y^i) > \tau^{rele} \right\},
\end{equation}
where $rele(\cdot)$ and $\tau^{rele}$ denote the relevance method and its threshold of \textit{<query, rewrite>} pair. The detail of the function $rele(\cdot)$ is discussed in the Section~\ref{sec:Feedback}.

Furthermore, Taobao's previous-generation of rewriting models primarily lacks optimization for long-tail queries.
As we work on the new generation of rewriting models, our goal is to maintain the relevance of retrievals while expanding the semantics.
This expansion is aimed at alleviating the issue of long-tail queries leading to \textit{``\nothing''} results. 
As a result, we utilize rejection sampling once again to filter $\mathcal{D}_{r}$ by considering the retrieval increment. 
Additionally, we include the most recent interacted product title of $x$ as supplementary information to better address long-tail queries: 

\begin{equation}
\begin{aligned}
\mathcal{D}_{sft}&=\{(concat(x^i, \mathcal{E}_x), y^i)|_{i=1} ^ {N_{sft}}  \\ &\text { such that } x^i, y^i \in \mathcal{D}_r, incr(x^i, y^i) > \tau^{incr} \}, 
\end{aligned}
\end{equation}
where, $\mathcal{E}_x$ is the interacted product title list of query $x$, $incr(\cdot)$ and $\tau^{incr}$ denote the increment method and its threshold of query-rewrite pair, respectively. The detail of the function $incr(\cdot)$ is discussed in the Section~\ref{sec:Feedback}.

\noindent \textbf{Auxiliary Task Datasets}: 
In order to further enhance LLMs' ability of comprehending long-tail queries, we have gathered three high-related task datasets in the context of query rewriting. These tasks encompass quality classification, product title prediction, and Chain-of-thought.
1) To tackle the quality classification task, our approach began with the extraction of query pairs from online logs. 
These query pairs were then subjected to human annotation to determine if they met the data requirements specified for SFT.
2) For the product title prediction task, we chose the most recent interacted product under the query as the reference, forming \textit{<query, product title>} pairs.
3) As for the CoT task, we employed the original online queries to construct prompts for human evaluators. 
It's noteworthy that these evaluators were not only tasked with providing query rewrites aimed at improving the quality of query retrieval but were also expected to articulate their thought processes, explaining the rationale behind their specific revisions. 
The details prompt design for above auxiliary tasks are shown in Table~\ref{tab:prompt-example}.
The collected data were later integrated into the rewriting task to form the final dataset, which was then randomly shuffled for the subsequent SFT stage.
\vspace{0.1cm}

\noindent \textbf{Supervised Fine Tuning}:
The process of generating text with a condition language model can be viewed as a constrained auto-regressive sampling strategy. Given a prompt $x$ and its gold standard $y$, 
The training objective is to maximize the conditional probability $p(y|x)$. Considering our multi-instruction SFT dataset and assuming that $p(y|x)=\prod_{i=1}{p(y_i|y_{0:i-1}, x)}$, the training objective of rewriting model involves minimizing the negative log likelihood:

\begin{equation}
\mathcal{L}_{\mathrm{SFT}}(\theta)=-\mathbb{E}_{(x, y) \sim \mathcal{D}_{msft}}\sum_{i=1} \log \pi\left(y_i \mid y_{0: i-1}, x; \theta\right),
\end{equation}
where $\mathcal{D}_{msft}$ denotes multi-instruction SFT data, which consists of a mixture of query rewriting dataset $\mathcal{D}_{sft}$ and a variety of auxiliary task datasets,  $\pi(\cdot)$ and $\theta$ denote our query rewriting model and its parameters. It is worth mentioning that LLMs typically have fixed prefixes in the prompt $x$. To avoid introducing noise, we disregarded the losses coressponding to $x$.

\begin{table}[]
\caption{Prompt examples of different instructions.}
\resizebox{0.42\textwidth}{!}{
\begin{tabular}{c|c}
\Xhline{1.5pt}
Task & Prompt Example \\ \midrule
\makecell{Quality \\ Classification} & \makecell[{{>{\parindent 0.2em}p{6cm}}}]{Is this a good e-commerce query rewrite? \\ Query: \{query\} \\ Rewrite: \{rewrite\} \\ System: \{Yes or No\}} \\ \midrule
Title Prediction &  \makecell[{{>{\parindent 0.2em}p{6cm}}}]{Please generate product titles that match \\ input query \\ Query: \{query\} \\ System: \{product title\}} \\ \midrule
\makecell{Chain \\ of \\ Thought} & \makecell[{{>{\parindent 0.2em}p{6cm}}}]{Your task is to rewrite the input query into a \\ query that makes it easier to search for related \\ products, and you are required to give the tho-\\-ught process and then the query rewriting re-\\-sult. The thought process and the query rewr-\\-iting result are separated by a semicolon. \\ Query: \{query\} \\ System: \{CoT\}; \{Rewrite\} }\\ \Xhline{1.5pt}
\end{tabular}
}
\label{tab:prompt-example}
\end{table}

\subsection{Offline Feedback}
\label{sec:Feedback}
Currently, most alignment methods \cite{gulcehre2023reinforced, rafailov2023direct, mohankumar2021diversity, Agrawal2023} rely on manual annotation and training-based reward models. 
However, we argue that these approaches can be easily influenced by the quality of annotations and the effectiveness of the reward model training, which often leads to inaccurate reflection of response scores and compromises the learning of the generation model. 
To address this issue, we propose a feedback system based on the Taobao search engine, which provides more accurate rewrite scores.

The main structure of the Taobao online service process is illustrated in Figure~\ref{fig:taobao-framework}. 
When a query is received by the Taobao search engine, it undergoes preprocessing and is then passed to the query understanding. 
This module comprehends the semantic meaning of the query and send it to the retrieval module. The retrieval module retrieves a large number of candidate products from a massive Taobao product dataset. 
After deduplication and filtering, the candidate product sets from different retrieval paths are merged into an unordered and non-repetitive product set.
These products are presented to the user through a complex ranking system.

Similarly, when our offline system receives a rewrite, it simulates the process of the Taobao online service to retrieve the corresponding products for the rewrite.
Based on the product set, our system provides us with a quality score. 
It is important to note that we mainly address the semantic gap issue caused by long-tail queries in exact match. 
Therefore, our rewriting module only operates on the inverted index matching of the retrieval module, and the product set considered for rewrite retrieval is related only to the inverted index path. 
We propose three scores to measure the quality of the rewrite, namely relevance, increment, and hitrate.

As mentioned in Section~\ref{sec:MSFT} , even if the query and rewrite are semantically similar, it does not guarantee that the retrieved product sets will also be related. 
To prevent the model from retrieving products that are completely different from the user's original intent, we introduce the relevance score. It is calculated as follows:

\begin{equation}
rele(x, y) = \frac{\sum_i\mathbbm{1}_{f(x, z^i_y)>\tau'}}{|\mathcal{Z}_y|}, z^i_y \in \mathcal{Z}_y,
\end{equation}
where $\mathbbm{1}$ is the indicator function, $f(\cdot)$ is Taobao relevance function which is design for evaluate the relevance between item title and query text, $\tau'$ denotes the semantic relevance threshold of query-item pair, $\mathcal{Z}_y$ denote the offline retrieval product list of text $y$. These notations represent the same meaning in all the following formulas.

Furthermore, we need to ensure that the rewrite can expand the semantic meaning of the original query to some extent, avoiding the issue of \textit{``\nothing.''}
Therefore, we introduce the increment score, calculated as follows:

\begin{equation}
incr(x, y) = \frac{\sum_i\mathbbm{1}_{f(x, z_{xy}^i)>\tau'}}{\sum_i\mathbbm{1}_{f(x, z_{x'}^i)>\tau'}}, z_{xy}^i \in \mathcal{Z}_e \cap (\mathcal{Z}_x \cup \mathcal{Z}_y), z_{x'}^i \in \mathcal{Z}_e \cap \mathcal{Z}_x,
\end{equation}
where, 
$\mathcal{Z}_x$ denotes the offline retrieval product set of text $x$, and $\mathcal{Z}_e$ is the excellent product set maintained by Taobao search group.

Lastly, we define the hitrate, which measures to what extent the rewrite can compensate the semantic gap of the original query. The calculation process is as follows:

\begin{equation}
hitrate(x, y) = \frac{|\mathcal{E} \cap (\mathcal{Z}_x \cup \mathcal{Z}_y)|}{\sum_i\mathbbm{1}_{f(x, e^i)>\tau'}}, e^i \in \mathcal{E},
\end{equation}
where $\mathcal{E}$ is the collection of products that users have transacted outside of the search scenario, $\mathcal{Z}_x$ and $\mathcal{Z}_y$ denote the offline retrieval product set of text $x$ and $y$. 
Assuming that a product, which is semantically related to the current user's query, is not transacted in the search. This indicates that the original query fails to retrieve the product. However, if the rewrite is able to return the product, then it implies that the product exists in the $\mathcal{Z}_x \cup \mathcal{Z}_y$. This proves that the rewrite can indeed compensate for the semantic gap of the original query. As a result, the hitrate can effectively reflect the model's ability to compensate for this semantic gap.

Overall, our proposed feedback system based on the Taobao search engine provides more accurate rewrite scores by considering relevance, increment, and hitrate. This helps improve the alignment process and ensures better learning of the generation model.

\subsection{Objective Alignment}
\label{sec:OA}
To avoid introducing bias through the reward model, we introduce the Preference Rank Optimization (PRO)~\cite{song2023preference}, based on the Bradley Terry Model. 
This method aims to enforce the model to learn the rewrite partial order provided by offline feedback. According to the Bradley Terry Model, the probability of selecting an policy should be proportional to its corresponding reward. 
Given the partial order: $y_1 \succ y_2$, the preference probability can be expressed as:
\begin{equation}
P_{BT} = \frac{\exp(r(y_1, x))}{\exp(r(y_1, x))+\exp(r(y_2, x))},
\end{equation}
where $r(\cdot)$ is the reward function, which is defined as the normalized log probability of the rewrite generated by the LLM in PRO.

PRO expands this pairwise partial order to a more general list-wise partial order. 
Additionally, a temperature is introduced in order to capture the significance of the ranking based on the reward. 
PRO loss is represented by the following equation:
\begin{equation}
\mathcal{L}_{PRO}(\theta) = -\mathbb{E}_{(\boldsymbol{x}, \boldsymbol{y}) \sim \mathcal{D}_{PRO}}\sum_{k=1}^{n-1} \log \frac{\exp(\frac{\pi_{PRO}(y_k|x;\theta)}{\mathcal{T}^k_k})}{\sum_{i=k}^n \exp(\frac{\pi_{PRO}(y_i|x;\theta)}{\mathcal{T}^i_k})},
\end{equation}
where $\mathcal{T}^i_k=1/(r(y_k)-r(y_i))$ and  $\mathcal{T}^k_k=\min_{i>k}(\mathcal{T}^i_k)$for ranking difference, $\mathcal{D}_{PRO}$ denotes the dataset for alignment, $\pi_{PRO}(\cdot)$ and $\theta$ denote the policy model and its parameters, $n$ denotes the number of candidate rewrites.
Additionally, a SFT loss is added on top of the PRO loss with weight $\lambda$ to maintain the ability of the model to generate normal outputs:
\begin{equation}
\mathcal{L}_{ALIGN} = \mathcal{L}_{SFT}+\lambda\mathcal{L}_{PRO}.
\end{equation}

\subsection{Online Serving}
Due to the constraints imposed by numerous parameters and the autoregressive prediction mode of large language models, it is almost impractical to deploy \method online and meet the latency requirements of the Taobao search system.
To address this issue, we utilize \method offline to perform rewriting inference on torso and tail queries, which are defined as queries with retrieval results containing less than 70\% related products. 
The rewrites generated from these queries are stored in an online key-value graph, enabling quick online response. 
When a long-tail query matches the key-value graph, the search engine retrieves the corresponding rewrite. 
Subsequently, both the query and rewrite are tokenized into terms and used as keywords for inverted index matching to obtain a set of related products.
The union of the query and rewrite retrieval sets forms the final candidate set of products for the ranking system. 
The offline inference of \method covers 27\% of the page views (PV) in Taobao's main search and has a minimal impact on the latency of the online retrieval system.

%% file: src/experiments.tex
\section{Experiments}
\subsection{Datasets}
\textbf{Training Dataset}: For the SFT training process, we extracted approximately 20 million records from the online rewriting logs on Taobao prior to September 2023 as the initial training data. 
To enhance the quality of the training data, we conducted two rounds of rejection sampling to obtain the query rewriting dataset, which consists of 419,806 pairs of \textit{<query, rewrite>}. 
Additionally, we included 155,662 manually rewriting data in the dataset to ensure that the SFT model's rewriting adheres to human preferences. 
Finally, we combined 50,000 samples each from quality classification, product title prediction and CoT task with the query rewriting dataset to construct the Multi-instruction SFT dataset.

For the objective alignment training process, we randomly selected 10,000 queries from the query rewriting dataset to generate candidates. 
For each query, we used the SFT Model to generate 5 candidate queries. 
These 50,000 rewrites were then scored using the Taobao offline system.
After removing any outliers, the alignment training dataset consisted of a total of 45,350 candidate rewrites.

\noindent{\textbf{Test Dataset}}: For the offline test set, we selected 14,981 queries from Taobao search logs to evaluate the model's performance. 
Among them, 50\% of the queries were randomly sampled proportionally to the actual queries. 
Furthermore, to assess the model's capability to rewrite long-tail queries, we sampled the remaining 50\% of the data from the long-tail queries.


\subsection{Evaluation}
\textbf{Offline Metrics}: We employ relevance~(rele), increment~(incr) and hitrate as offline metrics to valid the effectiveness of our method. These metrics are defined in Section \ref{sec:Feedback}.

\noindent{\textbf{Online Metrics}}: We introduce three key online metrics: pay amount, number of transactions, and number of unique IP of visitors, to evaluate the model's online performance. 
These metrics are represented in abbreviated form as follows:
\begin{equation}
\begin{aligned}
GMV &:= \#\text{pay amount}, \\
\#Trans &:= \#\text{transaction},\\
UV &:= \#\text{unique IP of visitors.}
\end{aligned}
\end{equation}

\subsection{Implementation Details}
Unless stated otherwise, the optimizer used for model training is AdamW. 
During the multi-instruction SFT stage, the model is trained for one epoch with the learning rate set to 1e-5. 
In the objective alignment stage, the model is trained for four epochs with the learning rate set to 1e-6. 
Additionally, the maximum length of the prompt for the rewriting task is set to 64, while for the rest of the tasks it is set to 256.

\subsection{Offline Experiments}

\subsubsection{Results of Different LLMs}
\begin{table}[!t]
\caption{Comparison of different LLMs trained on multi-instruction SFT dataset.}
\begin{tabular}{c|c|ccc}
\Xhline{1.5pt}
Method  & \#params & \textit{rele}~(\%) & \textit{incr}~(\%) & \textit{hitrate}~(\%) \\ \midrule
    ChatGLM        & 6b      & 63.5 & 143.4 & 15.88\\
    ChatGLM2.0     & 6b         & 63.2 & 105.0 & 14.39 \\
    Baichuan       & 7b      & 66.0 & 114.3 & 15.70\\
    Qwen      & 7b        & 62.6 & 133.2 & 15.63\\ \Xhline{1.5pt}
\end{tabular}
\label{table:base-model}
\end{table}

In this section, we present the comparison results of different LLMs as base models trained on the multi-instruction SFT dataset, as shown in Table \ref{table:base-model}. 
The base models include ChatGLM~\cite{du2022glm, zeng2022glm}, ChatGLM2.0~\cite{du2022glm, zeng2022glm}, Baichuan~\cite{website:Baichuan}, and Qwen~\cite{qwen2023report}. 
The variability in the advantages of these base models can be attributed to disparities in their pre-training data and model architectures. 
Notably, ChatGLM exhibits superior performance in both increment and hitrate compared to the other three models. 
However, due to copyright constraints, we have to utilize Qwen as the base model to fulfill our business objectives.
Furthermore, Qwen's pre-training incorporates e-commerce data, augmenting its comprehension of the e-commerce domain. 
This indicates a higher potential for enhancing Qwen's ability to rewrite long-tail queries.

\subsubsection{Ablation Study of Multi-instruction SFT data}
To investigate the impact of the multi-instruction SFT dataset on the rewriting ability of LLMs, an ablation study was conducted in this section. 
First, we randomly extracted query pairs from online logs, with a size similar to that of the multi-instruction SFT dataset, which were not subjected to rejection sampling and were not mixed with other task data. Then, we used the data to train the model and obtained Qwen w/o MI. 
As depicted in Table \ref{table:mi}, it is evident that the multi-instruction SFT dataset significantly enhances the model's performance in terms of relevance, increment, and hitrate. 
This improvement can be attributed to the substantial enhancement in the quality of query pairs after rejection sampling. 
Furthermore, auxiliary tasks closely associated with query rewriting, such as quality classification, product title prediction, CoT, etc., further augment the model's comprehension of query semantics, subsequently elevating the quality of the produced rewriting results. 

\begin{table}[!t]
\caption{Comparison of Qwen with/without auxiliary tasks. MI means multi-instruction}
\resizebox{0.4\textwidth}{!}{
\begin{tabular}{cccc}
\Xhline{1.5pt}
Method  & \textit{rele}~(\%) & \textit{incr}~(\%) & \textit{hitrate}~(\%) \\ \midrule
    Qwen w/o MI     & 61.4 & 109.6 & 14.58\\
    Qwen       & 62.6 & 133.2 & 15.63 \\ \Xhline{1.5pt}
\end{tabular}
}
\label{table:mi}
\end{table}

\subsubsection{Results of Different Contrast Number}
In Table~\ref{tab:contrast-num}, we present the impact of varying contrastive numbers of candidates on model performance during the objective alignment stage.
It can be observed that, as the number of candidates increases, the relevance shows a consistent improvement, while the increment and hitrate decrease. 
This outcome can be attributed to our modified PRO, where all candidates are treated as gold standard and SFT Loss is calculated for each of them.
Consequently, a larger candidate pool implies an increase in the SFT loss weights and a decrease in the partial order learning weights, improving relevance of generated rewrites. 
Furthermore, there exists a trade-off between relevance and increment, where an increase in one metric necessitates sacrificing the other, leading to a negative impact on the increment metric.
In addition, it is important to mention that hitrate can be regarded as the increment with weak relevance constraint, which explains its decrease in this context. 
Considering the balance of the three metrics, we select the model with contrast number of 4 to be best checkpoint for overall comparison.

\begin{table}[!t]
\caption{Comparison of \method with different number of contrast candidates.}
\resizebox{0.38\textwidth}{!}{
\begin{tabular}{cccc}
\Xhline{1.5pt}
Number  & \textit{rele}~(\%) & \textit{incr}~(\%) & \textit{hitrate}~(\%) \\ \midrule
    2              & 53.3 & 215.6 & 17.66\\
    3             & 56.6 & 205.4 & 17.36\\
    4              & 57.7 & 198.7 & 17.27\\
    5              & 58.8 & 190.3 & 17.21\\ \Xhline{1.5pt}
\end{tabular}
}
\label{tab:contrast-num}
\end{table}

\begin{table*}[]
\caption{Overall performance of \method with multiple baselines. The best results are in bold, and the second-best results are underlined. ``Top Queries'' are defined as queries with retrieval results containing more than 70\% related products; ``Torso Queries'' are defined as queries with retrieval results containing 10\%-70\% related products; ``Tail Queries'' are defined as queries with retrieval results containing less than 10\% related products. All the metrics are the larger the better.}
\resizebox{0.9\textwidth}{!}{
\begin{tabular}{crrrrrrrrrrrr}
\Xhline{1.5pt}
\multirow{2}{*}{Method}                               & \multicolumn{3}{c}{Top Queries}                                                       & \multicolumn{3}{c}{Torso Queries}                                                     & \multicolumn{3}{c}{Tail Queries}                                                      & \multicolumn{3}{c}{All Queries}                                         \\ \cline{2-13} 
                                                      & \textit{rele} & \textit{incr} & \textit{hitrate}               & \textit{rele} & \textit{incr} & \textit{hitrate}               & \textit{rele} & \textit{incr} & \textit{hitrate}               & \textit{rele} & \textit{incr} & \textit{hitrate} \\ \midrule
\multicolumn{1}{c|}{CLE-QR}                           & {\underline{73.4}}         & 90.0               & \multicolumn{1}{c|}{13.16}          & 24.7               & 36.4               & \multicolumn{1}{c|}{15.77}          & 10.0               & 17.2               & \multicolumn{1}{c|}{12.36}          & {\underline{69.6}}         & 90.0               & 12.95                 \\
\multicolumn{1}{c|}{BART}                             & 67.5               & 106.0              & \multicolumn{1}{c|}{13.26}          & 27.9               & 130.0              & \multicolumn{1}{c|}{17.70}          & 8.4                & 27.5               & \multicolumn{1}{c|}{13.59}          & 62.2               & 100.0              & 13.56                 \\
\multicolumn{1}{c|}{Q2D (ChatGPT)}                     & 71.7               & 47.3               & \multicolumn{1}{c|}{12.96}          & {\underline{35.3}}         & 66.7               & \multicolumn{1}{c|}{16.86}          & {\underline{12.8}}         & 15.9               & \multicolumn{1}{c|}{17.21}          & 66.7               & 45.5               & 14.73                 \\
\multicolumn{1}{c|}{Qwen (SFT)}                        & 67.1               & 117.4              & \multicolumn{1}{c|}{14.18}          & 25.5               & 56.4               & \multicolumn{1}{c|}{17.49}          & 7.2                & 34.9               & \multicolumn{1}{c|}{14.91}          & 61.4               & 109.6             & 14.58                 \\
\multicolumn{1}{c|}{RL (rele)}                        & \textbf{74.7}      & 48.6               & \multicolumn{1}{c|}{12.66}          & \textbf{39.4}      & 2.1                & \multicolumn{1}{c|}{15.60}          & \textbf{14.5}      & 8.1                & \multicolumn{1}{c|}{11.36}          & \textbf{70.0}      & 45.1               & 12.42                 \\
\multicolumn{1}{c|}{RL (incr)}                        & 46.9               & 76.1               & \multicolumn{1}{c|}{13.33}          & 17.7               & 134.1              & \multicolumn{1}{c|}{18.00}          & 4.4                & 60.9               & \multicolumn{1}{c|}{15.20}          & 42.5               & 75.9               & 14.22                 \\
\multicolumn{1}{c|}{RL (hitrate)}                     & 56.0               & 176.8              & \multicolumn{1}{c|}{15.10}          & 6.2                & 117.3              & \multicolumn{1}{c|}{{\underline{20.16}}}    & 2.2                & 51.0               & \multicolumn{1}{c|}{18.22}          & 49.6               & 162.8              & 15.75                 \\ \midrule
\multicolumn{1}{c|}{\method (rele)}    & 69.3               & 174.5              & \multicolumn{1}{c|}{15.04}          & 27.3               & 160.9              & \multicolumn{1}{c|}{19.29}          & 4.1                & {\underline{66.4}}         & \multicolumn{1}{c|}{{\underline{18.27}}}    & 62.3               & 164.0              & 16.43                 \\
\multicolumn{1}{c|}{\method (incr)}    & 64.6               & \textbf{212.3}     & \multicolumn{1}{c|}{\textbf{15.45}} & 20.5               & \textbf{207.7}     & \multicolumn{1}{c|}{\textbf{21.45}} & 2.4                & \textbf{71.7}      & \multicolumn{1}{c|}{\textbf{19.62}} & 57.7               & \textbf{198.7}     & \textbf{17.27}        \\
\multicolumn{1}{c|}{\method (hitrate)} & 69.3               & {\underline{177.0}}        & \multicolumn{1}{c|}{{\underline{15.39}}}    & 18.4               & {\underline{204.0}}        & \multicolumn{1}{c|}{19.78}          & 5.0                & 62.5               & \multicolumn{1}{c|}{18.22}          & 62.1               & {\underline{167.0}}        & {\underline{16.64}}  \\ \Xhline{1.5pt}        
\end{tabular}
}
\label{tab:main-results}
\end{table*}

\subsubsection{Main Results}
We compare \method with multiple baselines, including CLE-QR, query2doc~(Q2D), BART, Qwen, and RL-based LLM.
CLE-QR~\cite{li2022query} is the previous-generation query rewriter of Taobao search that generates semantic representations and retrieve related rewrites for each query based on contrastive learning.
BART~\cite{lewis2020bart} is a powerful pre-trained generation model based on the encoder-decoder structure. We fine-tune it with query pairs from online logs to enhance its ability to rewrite e-commerce queries.
Qwen~\cite{qwen2023report} is a large-scale language model based on the decoder-only structure that contains 7 Billion parameters. Similarly, we fine-tune it with query pairs from online logs to enhance its ability to rewrite e-commerce queries.
Furthermore, following the settings of \cite{Agrawal2023}, we introduce an RL-based LLM and utilize relevance, increment, and hitrate as rewards to encourage the RL model to align with the Taobao offline metrics, respectively.
From analyzing the data presented in Table \ref{tab:main-results}, the following conclusions can be drawn:

\textbf{Generative models outperform discriminative models when rewriting ``Torso'' and ``Tail'' queries.} 
For instance, when considering CLE-QR and BART, both models exhibit similar performance on ``Top Queries'' across three metrics.
However, BART significantly outperforms CLE-QR in terms of hitrate and increment on ``Torso Queries'' and ``Tail Queries'' while maintaining relevance. 
This discrepancy arises because discriminative models like CLE-QR rely on existing queries in the search system as rewrite candidates, which are often biased towards top queries. 
As a result, torso and tail queries, which lack semantically similar top rewrites, do not receive related search candidates from CLE-QR. 
In contrast, BART's rewriting process is not restricted by the semantic scope of online queries, enabling it to generate rewrites that are not present in the search system's history. 
This allows BART to overcome the limitations of discriminative models and optimize torso and tail query rewriting problem.

\textbf{LLMs exhibit superior long-tail semantic understanding capabilities compared to small models. }
Qwen and BART serve as examples, where Qwen, with its extensive parameter size, demonstrates stronger semantic expansion than BART in terms of hitrate and increment of ``All Queries''. 
Analyzing individual query slices, Qwen's improvement in hitrate and increment primarily occurs in the ``Tail Queries'', further validating the suitability of LLMs for long-tail query rewriting tasks.

\textbf{Retrieval augmentation methods demonstrate limited semantic expansion capabilities.}
Comparing Q2D~(ChatGPT) and \method, Q2D~(ChatGPT) maintains good retrieval relevance across all query slices but lacks sufficient semantic expansion capabilities, resulting in subpar increment and hitrate performance. 
Conversely, our \method, which is specifically optimized for semantic expansion in rewriting, significantly enhances these two metrics.

\textbf{The reinforcement learning~(RL) may introduce bias and impact the effectiveness of the rewriting LLMs.} 
Examining RL and \method, RL process introduces a reward model to guide the base model's training. 
However, calculating the reward requires offline search system simulation, and the reward model may not accurately capture the search system's features, leading to reduced performance of RL models. 
In contrast, our \method employs contrastive learning to explicitly learn the partial order of candidates, circumventing potential bias caused by the reward model. 
Ultimately, while minimizing the adverse impact on retrieval relevance, \method substantially improves the model's increment and hitrate.

\textbf{Different offline metrics work differently as rewards.} 
For instance, when the \method framework prioritizes relevance as its training objective, it demonstrates a more cautious approach to bridging the semantic gap.
The improvements in both increment and hitrate tend to be challenging to achieve in this context.
However, when the primary objective shifts to maximizing increment, the model demonstrates a significant capacity to enhance both the increment and hitrate of retrieval, effectively addressing the issue of \textit{``\nothing''}.
In such cases, a marginal decrease in relevance becomes an acceptable trade-off.
When hitrate becomes the target, the model also can effectively enhance both the increment and hitrate.
Nevertheless, owing to the intricacies of the hitrate computation process, the model encounters difficulties in capturing the partial order among candidates. 
Consequently, the model's ability to expand semantic is diminished in comparison to the \method that focuses on increment.

\subsection{Online Experiments}
\begin{table}[!t]
\caption{Online A/B test of \method on Mobile Taobao Search. 
``all queries'': every query in the test bucket counts, regardless of whether it has been rewritten or not.
``covered queries'': only rewritten queries count.
``long-tail queries'': rewritten and long-tail queries count.
``\nothing queries'': rewritten and \nothing queries count.}
\begin{tabular}{c|ccc}
\Xhline{1.5pt}
Online Traffic & GMV & \#Trans & UV \\ \midrule
    all queries   & +0.40\%  & +0.34\% &  +0.33\% \\
    covered queries    & +2.96\% & +1.36\% & +1.22\% \\
    long-tail queries &  +1.57\% & +2.52\% & +2.32\% \\
    \textit{``\nothing''} queries & +18.66\% & +5.90\%  & +6.25\%  \\ \Xhline{1.5pt}
\end{tabular}
\label{tab:online-results}
\end{table}

To assess the actual online performance of \method, we deployed it on Taobao search for a 14-day online test, during which we recorded the three key metrics in the Taobao search scene: GMV, \#Trans, and UV.
Table~\ref{tab:online-results} reveals that \method surpassed the previous-generation rewriting model CLE-QR by 0.4\%, 0.34\%, and 0.33\% in terms of GMV, \#Trans, and UV, respectively. This implies that \method contributes millions of GMV to Taobao search. 
It's important to note that the overall performance mentioned here refers to all queries in the testing buckets. 
Since we inference offline, there are about 70\% of online queries that do not hit our rewriting table. 
Even in these cases, our model still delivers remarkable enhancements.
Additionally, for the queries covered~(rewritten) by \method (approximately 27\% of total PV), there were noteworthy increases of 2.96\%, 1.36\%, and 1.22\% in GMV, \#Trans, and UV, respectively. 
These findings indicate that \method effectively rewrites queries and addresses potential semantic gaps in the semantic matching process.
Moreover, \method significantly improved online \#Trans and UV for long-tail queries and \textit{``\nothing''} queries, although we disregarded the GMV fluctuation for this subset due to its low proportion. 
This improvement can be attributed to our specialized optimization for long-tail queries. 
During the first-stage SFT of \method, rejection sampling and auxiliary task data enhanced the model's performance in terms of retrieval increment and relevance, and also deepened its understanding of long-tail queries. 
The alignment process in the second and third stages effectively compelled the model to align with online objectives of Taobao search.

%% file: src/conclusion.tex
\section{Conclusion}
In this paper, we introduce \method, a framework specifically designed for e-commerce query rewriting. The main objective of \method is to address the semantic gap that occurs during the semantic matching process, particularly for long-tail queries. Initially, we improve the quality of the rewriting dataset by employing rejection sampling and auxiliary task mixing. We then train a LLM using this refined dataset, which enhances the model's query understanding and enables effective rewriting of e-commerce queries.
Utilizing the well-trained LLM, we generate multiple candidate rewrites for each sampled query. To establish a partial order among these candidates, we create an offline feedback system based on online Taobao search. This feedback system accurately evaluates the retrieval quality of the candidate rewrites from various perspectives, such as relevance, increment, and hitrate. Finally, by incorporating the partial order of rewriting retrieval quality, we introduce PRO, which aligns the model's objectives with those of Taobao. This ensures that our approach generates rewriting results that yield high-quality retrieval outcomes.
Through multiple experiments, we have demonstrated the effectiveness of our approach in improving offline metrics. Additionally, online A/B experiments have substantiated a significant increase in Taobao Search's GMV, \#Trans, and UV, particularly for long-tail queries.

\section{Acknowledgments.} 
This work was supported in part by the grants from National Natural Science Foundation of China (No.62222213, U22B2059, 62072423), and the USTC Research Funds of the Double First-Class Initiative (No.YD2150002009).